\documentclass[preprint,
showpacs,
amsmath,amssymb,
aps
]{revtex4-1}
\usepackage{dcolumn}% Align table columns on decimal point
\usepackage{bm}% bold math
\usepackage{caption}
\usepackage{mathrsfs}
\usepackage[titletoc]{appendix}
\usepackage{graphicx}
\usepackage{epstopdf}
\usepackage{hyperref}
\usepackage{subfigure}
\usepackage{slashed}

\begin{document}
	
	\title{Gauge dependence of spontaneous radiation  spectrum in a time-dependent relativistic non-perturbative Coulomb field}
	\author{Xue-Nan Chen}
	\author{Yu-Hang Luo}
	\author{Xiang-Song Chen}
	\email{For correspondence: cxs@hust.edu.cn}
	
	\affiliation{Hubei Key Laboratory of Gravitational and Quantum Physics, School of Physics, Huazhong University of Science and Technology, Wuhan 430074, China}
	
	\date{\today}
	\begin{abstract}
	The delicacy of gauge choice in calculating atomic transitions was first raised by Lamb and gave arise to intensive discussion as well as much controversy. These discussion and controversy focused on choosing a proper gauge for the electromagnetic wave that interacts with an atom. The issue was claimed to have been solved, especially by Lamb himself and co-workers, by favoring a gauge-invariant Hamiltonian for defining the atomic  state in the presence of electromagnetic wave. 
	Here we extend the problem to include a time-dependent relativistic non-perturbative Coulomb field,  which can be produced by a cluster of relativistic charged particles. If adiabatic conditions are carefully maintained, such a field must be included along side the nuclear Coulomb potential when defining the atomic state. We reveal that when taking the external field approximation, the gauge choice for this time-dependent relativistic non-perturbative Coulomb field cannot be overcome by previous method, and leads to considerable gauge-dependence of the transient spontaneous radiation spectrum. We calculate explicitly with a simple one-dimensional charged harmonic oscillator that such a gauge-dependence can be of a measurable magnitude of 10 MHz or larger for the commonly used Coulomb, Lorentz, and multipolar gauges. 
	Contrary to the popular view, we explain that this gauge dependence is not really a disaster, but actually an advantage here: The relativistic bound-state problem is so complicated that a fully quantum-field method is still lacking, thus the external field approximation cannot be derived and hence not guaranteed. However, by fitting to the experimental data, one may always define an {\it effective} external field, which may likely be parameterized with the gauge potential in a particular gauge. This effective external field would not only be of phenomenological use, but also shed light on the physical significance of the gauge field. 
		
	\end{abstract}

	\maketitle
	
	\section{Introduction}
	Gauge symmetry is trivial in classical electrodynamics, which is essentially a theory of the electric field $\vec E$ and magnetic field $\vec B$. A quantum formulation of electromagnetic interaction has to utilize the gauge potential $A^\mu$,  therefore does not share the absolutely safe gauge-invariance as in the classical case. It is known that for scattering problems of elementary particles, gauge symmetry is still well preserved in a quantum theory. However, the issue seems to be tricky as a bound state is involved.  It was first noticed by Lamb about 70 years ago \cite{Lamb52} when studying the renowned Lamb-shift transition $2S_{1/2}\to 2P_{1/2}$ that a straightforward calculation can correctly give the observed line-shape in just one particular gauge, namely the so-called length gauge with the $-q\vec E \cdot \vec r$ interaction.  Here, ``straightforward  calculation'' means that the atomic state is defined as the eigenstate of the conventional energy operator 
	\begin{equation}{\label{E0}}
		{\cal E}_0=\frac{{\vec p}^2}{2m}+qV(\vec x),~~{\cal E}_0\left| \psi_n^0 \right>=E_n\left| \psi_n ^0\right>
	\end{equation}
	where $n$ labels the state, $\vec p=-i\vec \nabla$ (we set $\hbar=c=1$) is the canonical momentum operator, $q=-e$ is the electron charge,  and $V(\vec x)$ is the nuclear Coulomb potential.
	
	Lamb's observation is somehow surprising, and also puzzling, because the quantum-mechanical formulation of electromagnetic interaction, for example, in the Schr\"odinger equation
	\begin{equation}{\label{Schro}}
		i\partial _t \psi (\vec x, t)=H(t)\psi=
		\left\{\frac{1}{2m}(\vec p-q\vec A(\vec x, t))^2+q\phi(\vec x,t)+qV(\vec x)\right\} \psi(\vec x,t),
	\end{equation}
	is constructed {\em purposely} to be invariant under the joint gauge transformation 
	\begin{equation}{\label{trans}}
		\psi = U\psi ' =e^{-iq\Lambda (\vec x,t)} \psi ' ,
		~A_\mu = A_\mu ' -\frac iq U\partial_\mu U^{-1}
		= A_\mu '+\partial_\mu \Lambda(\vec x,t),
	\end{equation}
	where $U=e^{-iq\Lambda (\vec x,t)}$, with $\Lambda (\vec x,t)$ an arbitrary function.
	
	At the root of  Lamb's observation is that although the Schr\"odinger equation is gauge-invariant,  the energy operator ${\cal E}_0$ in Eq. (\ref{E0}) has a gauge-dependent expectation value.  A way out of this difficulty is therefore quite natural, as many authors argued \cite{Yang76,Yang81,Au84,Lamb87,Funai19}, that in the presence of an electromagnetic field the atomic state should be defined instead as the eigenstate of the gauge-invariant energy operator 
	\begin{equation}{\label{EA}}
		{\cal E}_A(t)=\frac{{\vec \pi}^2}{2m}+qV(\vec x),~~{\cal E}_A(t)\left| \psi_n^A (t)\right>=E_n\left| \psi_n^A(t) \right>,
	\end{equation}
	where $\vec \pi= \vec p-q\vec A(\vec x, t)$ is the mechanical momentum operator. Accidentally, in the length gauge $\vec A=0$, ${\cal E}_A$ coincides with ${\cal E}_0$, thus the ``straightforward  calculation'' can work.  As Lamb {\it et. al.} elaborated in Ref. \cite{Lamb87}, a gauge with non-zero $\vec A$ leads to complication, but careful and consistent calculations can give the same result as in the length gauge. The major cause of complication is that the operator ${\cal E}_A(t)$ is time-dependent, and normally does not commute at different times, therefore Eq. (\ref{EA}) defines instantaneous instead of stationary eigenstates $\psi_n^A(t) $, which are also time-dependent. (The eigenvalues $E_n$, nevertheless, are gauge-invariant and time-independent. Namely, $E_n$ do not depend on the explicit form of $\vec A$. We will come back to this simple but delicate point in the next Section.) 
	
	So far so good. The gauge-choice problem as Lamb raised was thus claimed to have been solved, especially by Lamb himself and co-workers in a ``concluding paper'' in 1987 \cite{Lamb87}.  
	However, we would like to remind that Eq. (\ref{EA}) cannot really be taken for granted, and is rather a {\em conjecture}.  Especially, what is the justification for Eq. (\ref{EA}) to include $qV(\vec x)$ but discard $q\phi(\vec x, t)$? Why not just take the total $H(t)$ in Eq. (\ref{Schro}) to define the instantaneous atomic eigenstates in the presence of electromagnetic interaction? \cite{Note} Certainly, the total $H(t)$ is gauge-dependent, and if Eq. (\ref{EA}) is replaced by 
	\begin{equation}{\label{Ht}}
		H(t)\left| \psi_n (t)\right>=E_n(t)\left| \psi_n(t) \right>,
	\end{equation}
	this would make both the eigenvalues and the atomic transition rates gauge-dependent. But quite interestingly, by Eq. (\ref{Ht}) the correct result can still be obtained in one particular gauge, namely the Coulomb gauge with $\phi=0$, instead of the length gauge with $\phi=-\vec E\cdot \vec x$. This time it is the Coulomb gauge that stands out.
	
	The aim of this paper to extend the previous studies, and discuss a more serious case that in Eq. (\ref{Schro}) the time-dependent scalar potential $\phi(\vec x,t)$ can be adiabatic and comparable to $V(\vec x)$ in effect, therefore must be treated at the same footing as $V(\vec x)$. Then, Eq. (\ref{EA}) could not possibly apply, and Eq. (\ref{Ht}) is a more reasonable option. This would give rise to significant gauge-dependence that could not be overcome by existing methods. The paper is organized as follows: In Section II, we introduce our physical system: a cluster of relativistic charged particles passing by a one-dimensional oscillator. Parameters can be adjusted to make  $\phi(\vec x,t)$ of the moving charge cluster to be adiabatic and non-perturbative for the oscillator, whose instantaneous eigenstates have to be calculated with  Eq. (\ref{Ht}) instead of Eq. (\ref{EA}).  The solutions differ significantly for the commonly used Coulomb, Lorentz, and multipolar gauges. Then in Section III we compute explicitly the transient spontaneous radiation spectrum of such a system, and find again significant gauge-dependence.  In the last Section we summarize our results and discuss their physical implications.

	\section{A time-dependent relativistic non-perturbative Coulomb field acting adiabatically on a quantum oscillator}
	
\begin{figure}[htbp]
		\centering
		\includegraphics[width=0.7\linewidth]{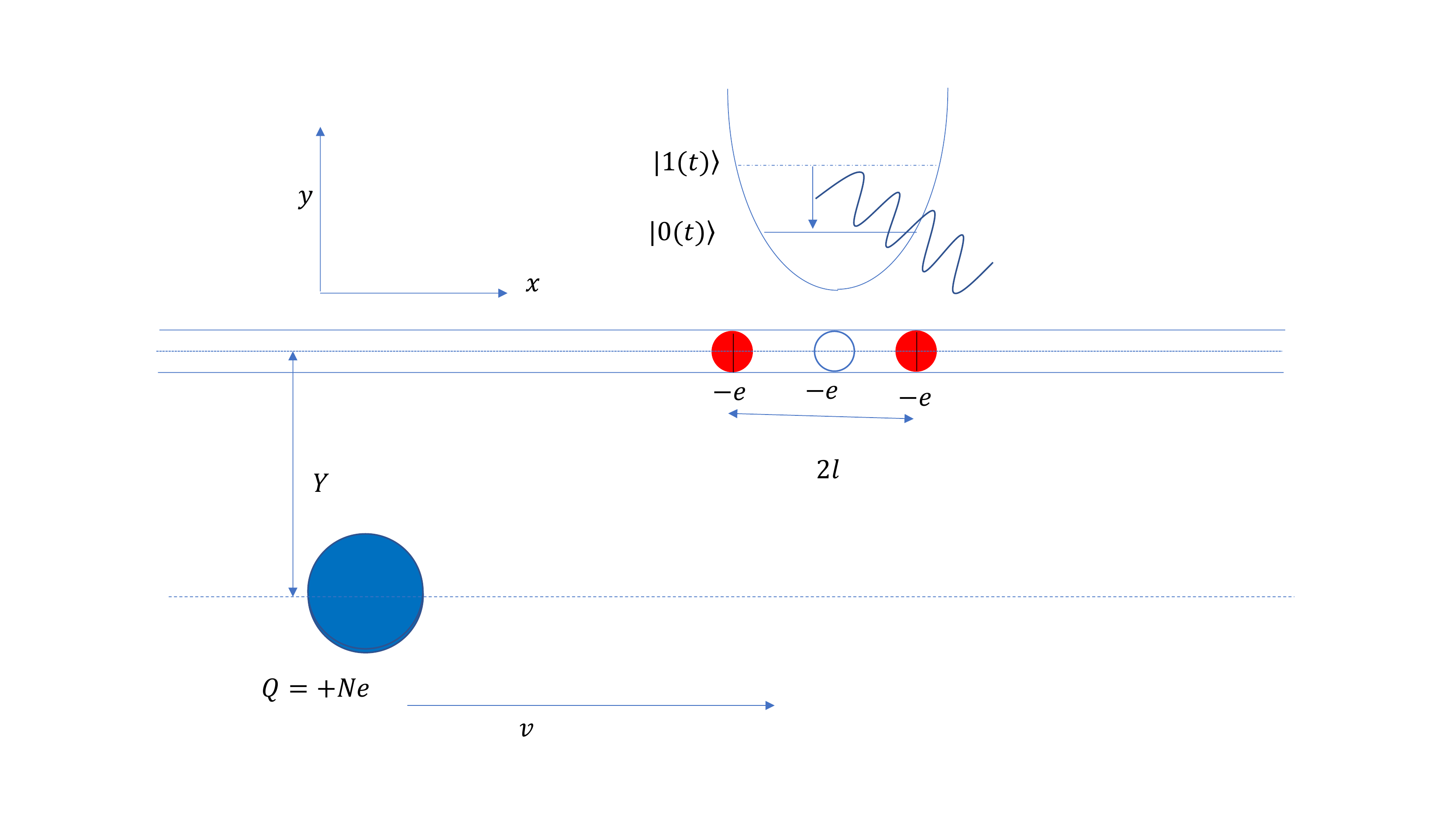}
		\captionsetup{justification=raggedright}
		\caption{The schematic design of the physical system. The big solid circle represents the proton cluster. Two small solid circles represent stationay charges. The small hollow circle represents the electron. See the text for description of parameters.}
		\label{fig:scheme2}
	\end{figure}

	FIG. \ref{fig:scheme2} shows a schematic design of our physical system. The electromagnetic field is produced by a cluster of relativistic protons of a huge number $N$, like a bunch from an accelerator. (Certainly electrons and heavy ions may also be utilized.)  This charge cluster passes by a one-dimensional oscillator, formed of an electron moving in a nanowire or carbon nanotube along the $x$ axis, and constrained by two other nearby negative charges, placed at the coordinates $(x,y)=(l,0)$ and $(-l,0)$, respectively. The charge cluster moves in the same direction, with an impact distance of $Y$. If excited, the oscillator can emit a photon by spontaneous radiation. It can be expected that by enlarging the parameters $N$, $l$, and $Y$,  the scalar potential $\phi(\vec x,t)$ of the charge cluster can be non-perturbative and adiabatic for the oscillator for a duration which is sufficiently long for the excited oscillator to emit a transient photon. This emission must then be computed by including both  $\phi(\vec x,t)$ and $V(\vec x)$ when defining the instantaneous eigenstates of the oscillator.  Namely, we have to employ Eq. (\ref{Ht}) instead of Eq. (\ref{EA}), as we remarked above in Section I.

	For simplicity, we treat the proton cluster as a point charge, and leave the task of considering the actual spatial distribution and possible dispersing of the cluster to a future study. This would influence some quantitative detail but not the main concern of this paper about gauge-dependence.  The cluster velocity $\beta=v/c$ should be chosen large enough to produce a considerable difference for $\phi (\vec x, t)$ among various gauges, and at the same time small enough to allow for a rough external-field approximation. (This point will be commented on further in the last Section.)
	
	Upon solving Eq. (\ref{Ht}) for our one-dimensional oscillator, we first note that it shares the same feature as Eq. (\ref{EA}) that the eigenvalues do not depend explicitly on $\vec A$. This is a simple fact, but may often cause confusion, so we elaborate a little bit here. 
	
	We may always apply to Eq. (\ref{Ht}) the following unitary transformation:
	\begin{equation}{\label{tilde}}
		\left| \psi_n(t) \right> = U \left| \tilde \psi_n(t) \right>,~H(t)=U^{-1} \tilde H(t) U, 
	\end{equation}
	then Eq. (\ref{Ht}) becomes
	\begin{equation}{\label{tildeHt}}
		\tilde H(t)\left|  \tilde \psi_n (t)\right>=E_n(t)\left|  \tilde \psi_n(t) \right>.
	\end{equation}
	
	Note that although the factor $U$ in Eq. (\ref{tilde}) can be the same as in Eq. (\ref{trans}), Eq. (\ref{tilde}) is {\em not} a gauge transformation. It is just a mathematical technique, after {\em fixing} a gauge, for the convenience of solving the eigen-equation.  Especially, the new operator 
	\begin{equation}\label{tildeH}
		\begin{aligned}
			\tilde H(t)=U H(t) U^{-1}&=\frac{1}{2m}(\vec p-q\vec A(\vec x, t)
			+q\vec \nabla \Lambda(\vec x,t ))^2
			+q\phi(\vec x,t)+qV(\vec x) \\
			&=\frac{1}{2m}(\vec p-q\vec A '(\vec x, t))^2+q\phi(\vec x,t)+qV(\vec x),
		\end{aligned}
	\end{equation}
	is just a mathematical tool without much physical meaning. It is {\em not} the gauge-transformed Hamiltonian $H'(t)$. The latter should be obtained by applying $\psi=U\psi '$ to the time-evolution equation (\ref{Schro}) instead of the eigen-equation (\ref{Ht}). The result is familiar: 
	\begin{equation}
		H'(t)=U H(t) U^{-1} -iU^{-1}\partial_t U=\frac{1}{2m}(\vec p-q\vec A '(\vec x, t))^2+q\phi '(\vec x,t)+qV(\vec x),
	\end{equation}
	where both the vector and scalar potentials are the gauge-transformed ones. In contrast, $\tilde H(t)$ looks like a hybrid-gauge expression: the vector potential is transformed, while the scalar potential remains the same.  
	
	It is exactly the hybrid feature of $\tilde H$ in Eq. (\ref{tildeH}) that makes the unitary transformation (\ref{tilde}) advantageous when solving Eq. (\ref{Ht}): For a one-dimensional problem, the vector potential $\vec A '$ may always be set to zero, and $\tilde H$ got simplified. After obtaining the simpler solution $\tilde \psi_n$, the original $\psi_n$ is easily got by multiplying the factor $U$, and the eigenvalue $E_n(t)$ is unchanged. Hence, the vector potential $\vec A$ is trivial when solving Eq. (\ref{Ht}), as we just commented above, and we need only to consider the scalar potential $\phi (\vec x, t)$.
	
	Since in our system the moving cluster and the fixed charges have comparable effects on the electron, we include all their contributions into $\phi (\vec x, t)$. The expression is easy to calculate. For the Lorentz gauge, 
	\begin{equation}
		\phi_{L}=\frac{1}{4 \pi}\left(\frac{-e}{l+x}+\frac{-e}{l-x}+\frac{N e}{\sqrt{(x-L(t))^{2}+\left(1-\beta^{2}\right) Y^{2}}}\right), 
	\end{equation}
	and for Coulomb gauge,
	\begin{equation}
		\phi_{C}=\frac{1}{4 \pi}\left(\frac{-e}{l+x}+\frac{-e}{l-x}+\frac{N e}{\sqrt{(x-L(t))^{2}+Y^{2}}}\right).
	\end{equation}
	Here, $x$ is the coordinate of the electron, $L(t)$ is the coordinate of the charge cluster at time $t$. The subscripts $L,C$ refer to expressions in the Lorentz and Coulomb gauges, respectively, and a subscript $G$ will denote a general gauge.  
	
	By applying the PZW transformation \cite{Power59,Wolley71}:
	\begin{equation}\label{PZW}
		U_G=\exp \left[i e \int_{0}^{1} d \lambda \vec{A}_G\left(\lambda \vec{x}+(1-\lambda) \vec{x}_{0}, t\right) \cdot\left(\vec{x}-\vec{x}_{0}\right)\right],
	\end{equation}
	one obtains  $\phi_{M}$ in the multipolar gauge, 
	\begin{equation}
		\phi_{M}=\left\{\phi_G(\vec x_0,t)-\int_{0}^{1} d \lambda \vec{E}\left(\lambda \vec{x}+(1-\lambda) \vec{x}_{0}, t\right) \cdot\left(\vec{x}-\vec{x}_{0}\right)\right\}.
	\end{equation}
	
	To further simplify our calculation, we approximate the electron motion as a harmonic oscillator, with the equilibrium point influenced by the charge cluster. To this end, we Taylor-expand $\phi_G$ at $x_0$ to the second order:
	\begin{equation}
		\phi_G(x,t)\approx\phi_G(x_0,t)+\partial_x \phi_G(x,t)|_{x=x_0}\cdot (x-x_0)+\frac 12 \partial_x^2 \phi_G(x,t)|_{x=x_0}\cdot (x-x_0)^2,
	\end{equation}
	then the value of  $x_0$ is found by solving $\partial_x \phi_G(x,t)|_{x=x_0}=0$. For the Lorentz gauge,
	\begin{equation}
		\begin{aligned}
			\phi_{L} &\approx\frac{1}{4 \pi}\left\{\frac{-e}{x_{0}+l}-\frac{-e}{x_{0}-l}+\frac{N e}{\sqrt{\left(x_{0}-L(t)\right)^{2}+\left(1-\beta^{2}\right) Y^{2}}}\right.\\
			&-\left(\frac{-e}{\left(x_{0}+l\right)^{2}}-\frac{-e}{\left(x_{0}-l\right)^{2}}+\frac{N e\left(x_{0}-L(t)\right)}{\sqrt{\left(x_{0}-L(t)\right)^{2}+\left(1-\beta^{2}\right) Y^{2}}^{3}}\right)\left(x-x_{0}\right) \\
			&+\left(\frac{-e}{\left(x_{0}+l\right)^{3}}-\frac{-e}{\left(x_{0}-l\right)^{3}}\right.\\
			&\left.\left.-\frac{1}{2}\left(\frac{N e}{\sqrt{\left(x_{0}-L(t)\right)^{2}+\left(1-\beta^{2}\right) Y^{2}}^{3}}-\frac{3 N e\left(x_{0}-L(t)\right)^{2}}{\sqrt{\left(x_{0}-L(t)\right)^{2}+\left(1-\beta^{2}\right) Y^{2}}^5}\right)\right)\left(x-x_{0}\right)^{2}\right\}.
		\end{aligned}
	\end{equation}
	For the Coulomb gauge
	\begin{equation}
		\begin{aligned}
			\phi_{C} &\approx\frac{1}{4 \pi}\left\{\frac{-e}{x_{0}+l}-\frac{-e}{x_{0}-l}+\frac{N e}{\sqrt{\left(x_{0}-L(t)\right)^{2}+Y^{2}}}\right.\\
			&-\left(\frac{-e}{\left(x_{0}+l\right)^{2}}-\frac{-e}{\left(x_{0}-l\right)^{2}}+\frac{N e\left(x_{0}-L(t)\right)}{\sqrt{\left(x_{0}-L(t)\right)^{2}+Y^{2}}^{3}}\right)\left(x-x_{0}\right) \\
			&\left.+\left(\frac{-e}{\left(x_{0}+l\right)^{3}}-\frac{-e}{\left(x_{0}-l\right)^{3}}-\frac{1}{2}\left(\frac{N e}{\sqrt{\left(x_{0}-L(t)\right)^{2}+Y^{2}}^{3}}-\frac{3 N e\left(x_{0}-L(t)\right)^{2}}{\sqrt{\left(x_{0}-L(t)\right)^{2}+Y^{2}}^{5}}\right)\right)\left(x-x_{0}\right)^{2}\right\}.
		\end{aligned}
	\end{equation}
	And for the multipolar gauge,
	\begin{equation}
		\begin{aligned}
			\phi_{M} &\approx\phi_G(\vec x_0, t)+\frac{1}{4\pi}\left\{-\left(\frac{-e}{\left(x_{0}+l\right)^{2}}-\frac{-e}{\left(x_{0}-l\right)^{2}}+\frac{N e\left(1-\beta^{2}\right)\left(x_{0}-L(t)\right)}{\sqrt{\left(x_{0}-L(t)\right)^{2}+\left(1-\beta^{2}\right) Y^{2}}^{3}}\right)\left(x-x_{0}\right)\right.\\
			&+\left(\frac{-e}{\left(x_{0}+l\right)^{3}}-\frac{-e}{\left(x_{0}-l\right)^{3}}\right.\\
			&\left.\left.-\frac{1}{2}\left(\frac{N e\left(1-\beta^{2}\right)}{\sqrt{\left(x_{0}-L(t)\right)^{2}+\left(1-\beta^{2}\right) Y^{2}}^{3}}-\frac{3 N e\left(1-\beta^{2}\right)\left(x_{0}-L(t)\right)^{2}}{\sqrt{\left(x_{0}-L(t)\right)^{2}+\left(1-\beta^{2}\right) Y^{2}}^{5}}\right)\right)\left(x-x_{0}\right)^{2}\right\}.
		\end{aligned}
	\end{equation}
	
	From the above expressions, it is clear that for a time-dependent electromagnetic field the so-obtained equilibrium point $x_0$ is gauge-dependent, and generally does not coincide with the point where $\vec E=0$, except in the multipolar gauge with $\vec A=0$. We therefore add the gauge label, and write the Hamiltonian of the oscillator approximately as 
	\begin{equation}\label{H0G}
		H_{0G}(t)\approx\frac {1}{2m}(p-i\partial_x U^\dagger_G)^2+\frac 12 k_G(t)(x-x_{0G}(t))^2.
	\end{equation}
	Here, \(k_G(t) \) is the strength of the harmonic potential, and varies with the equilibrium point. $U_G$ is the PZW-transformation factor given by Eq. (\ref{PZW}). For the multipolar gauge $U_G$  is simply unity. With this Hamiltonian (\ref{H0G}), we can solve the instantaneous eigenvalues and the instantaneous quantum states as
	\begin{equation}\label{nG}
		\begin{aligned}
			E_{n,G}(t)&=(n+\frac 12)\omega_G(t),\\
			|n_G(t)\rangle&=U^{\dagger}_GN_{nG}(t)\exp[-\frac 12\gamma_G(t)^2(x-x_{0G}(t))^2]\mathrm{H}_n [\gamma_G(t)(x-x_{0G}(t))].
		\end{aligned}
	\end{equation}
	Here, \(\omega_G(t)=\sqrt{k_G(t)/m}\), \(\gamma_G(t)=\sqrt{m\omega_G(t)}\), \(N_{n G}(t)=\left[\gamma_{G}(t) / \sqrt{\pi} 2^{n} n !\right]^{1/2}\), and \(\mathrm{H}_{n}\) is Hermite polynomial.
	
	In what follows, we set up the parameters and compute the numerical results. We take the physical mass of the electron, and keep in mind that its effective mass might be different in an actual system. After some rough estimation, we find that the following values suffice our study:  \(N=10^{12}\), \(\beta=0.1\),  \(l=6.33\mathrm{nm}\), and \(Y/l=10^6\). The cluster moves from $x=-100Y$ to $x=100Y$. 
	
\begin{figure}[htbp]
		\centering
		\subfigure[$x_{0M}$]{
			\includegraphics[width=10cm]{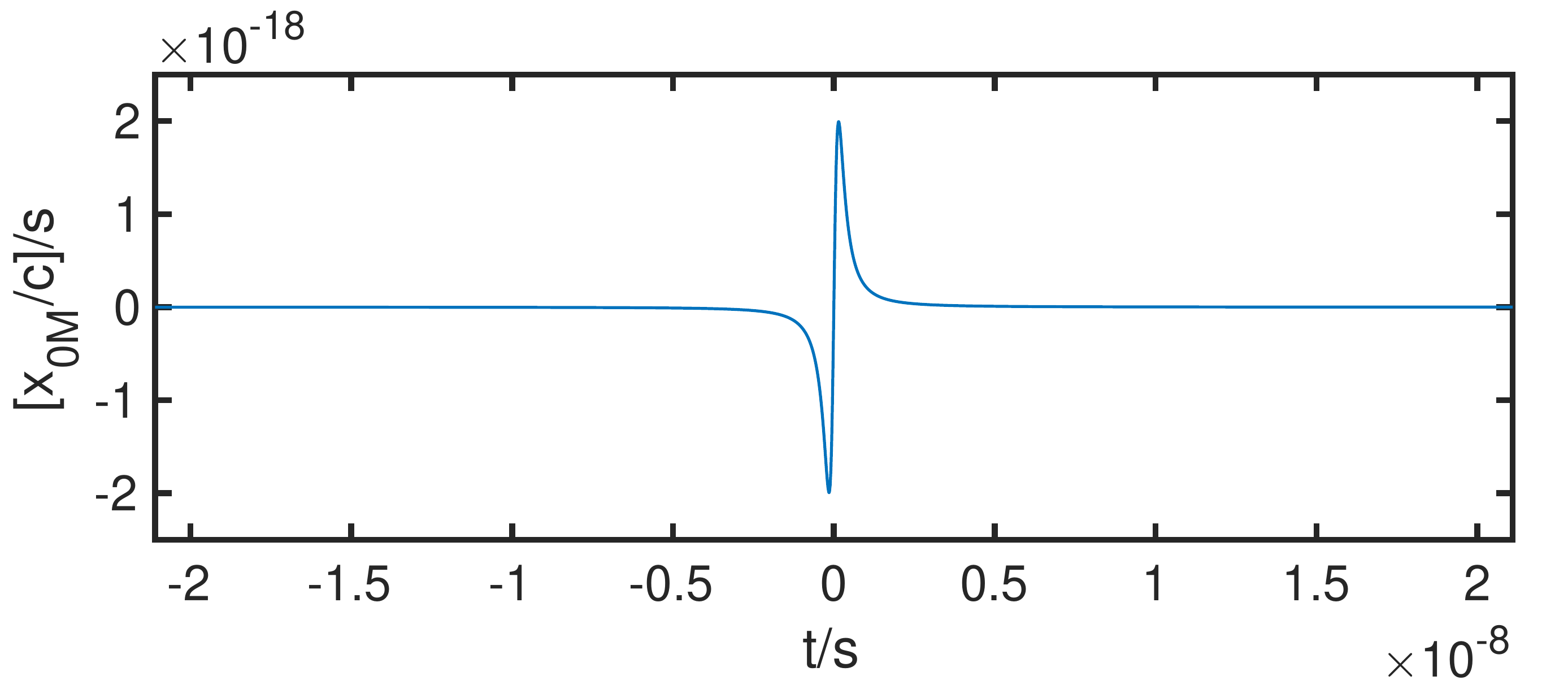}
		}
		\quad
		\subfigure[$\omega_{M}$]{
			\includegraphics[width=10cm]{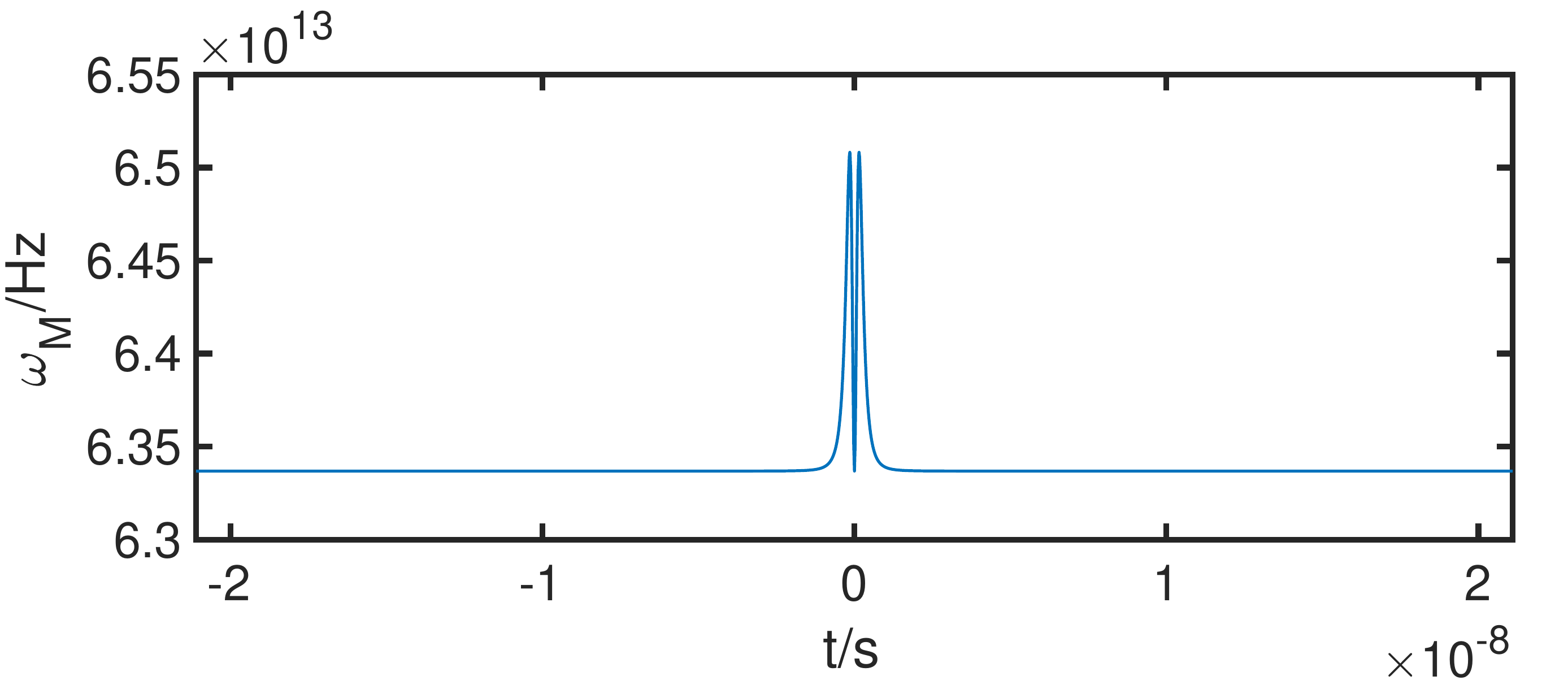}
		}
		\captionsetup{justification=raggedright}
		\caption{The result of the instantaneous equilibrium position (a) and the frequency (b) of the oscillator in multipolar gauge. To visualize the adiabatic chacacter of the system, we divide the equilibrium coordinate by $c$, and plot it in unit of time. }
		\label{fig:M}
	\end{figure}

\begin{figure}[htbp]
		\centering
		\subfigure[$x_{0M}$]{
			\includegraphics[width=10cm]{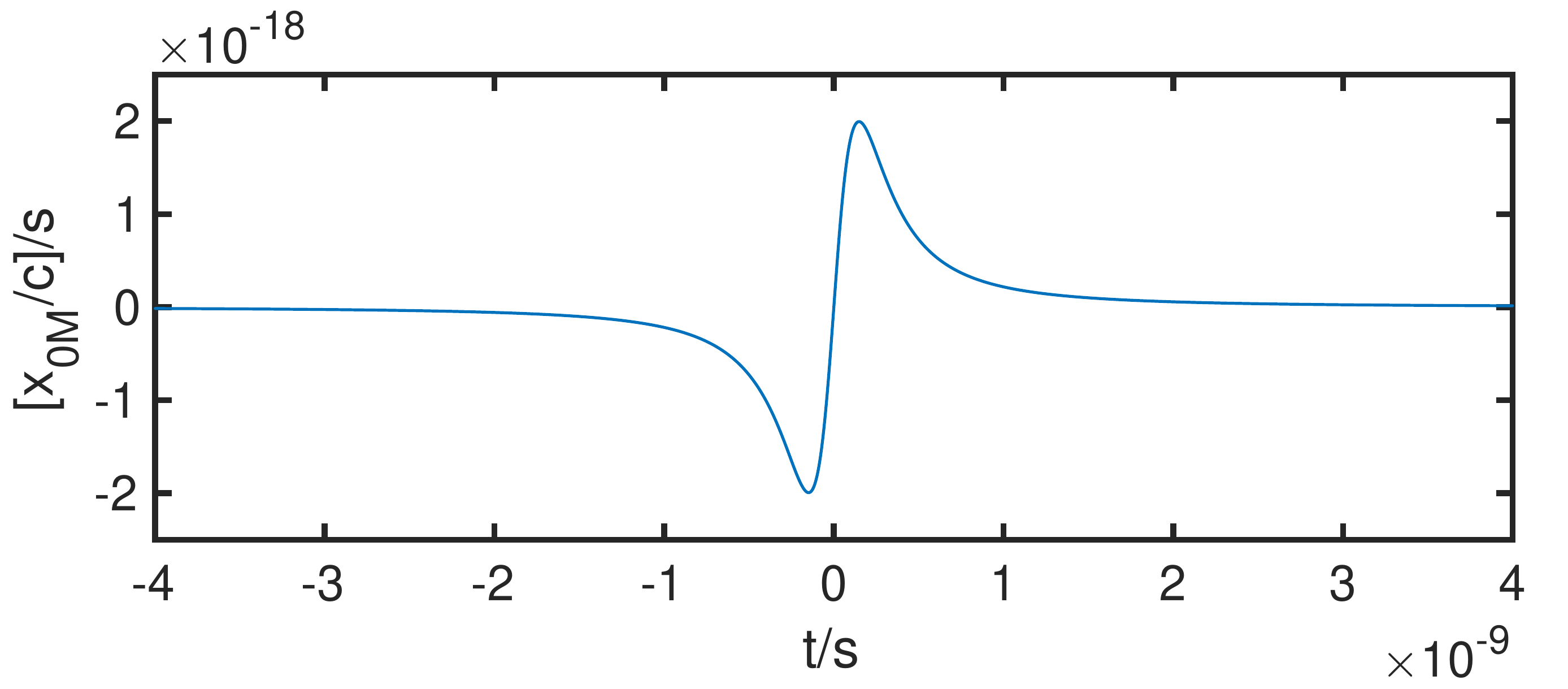}
		}
		\quad
		\subfigure[$\omega_{M}$]{
			\includegraphics[width=10cm]{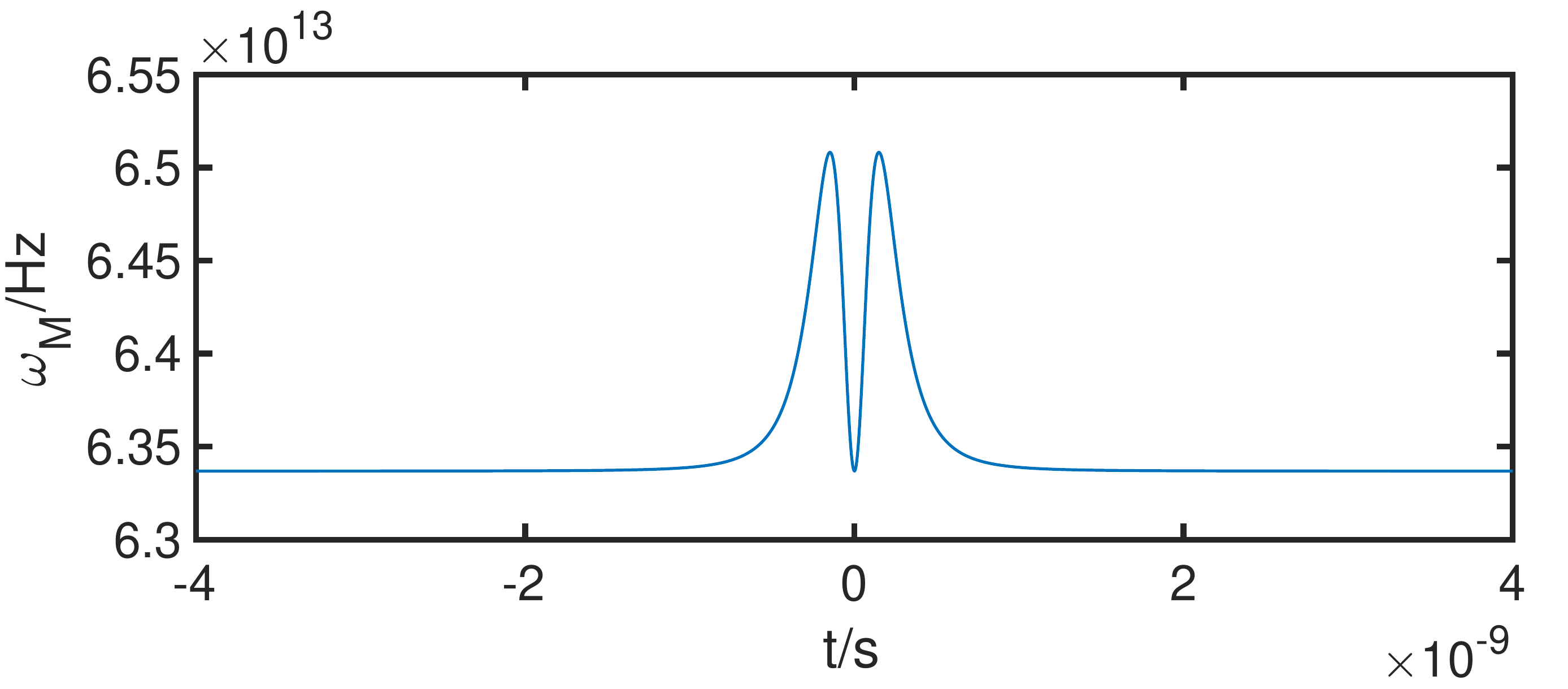}
		}
		\captionsetup{justification=raggedright}
		\caption{The enlarged view of FIG. \ref{fig:M} for the period when the cluster acts significantly.}
		\label{fig:M1}
	\end{figure}

\begin{figure}[htbp]
		\centering
		\subfigure[$x_{0G}-x_{0M}$]{
			\includegraphics[width=10cm]{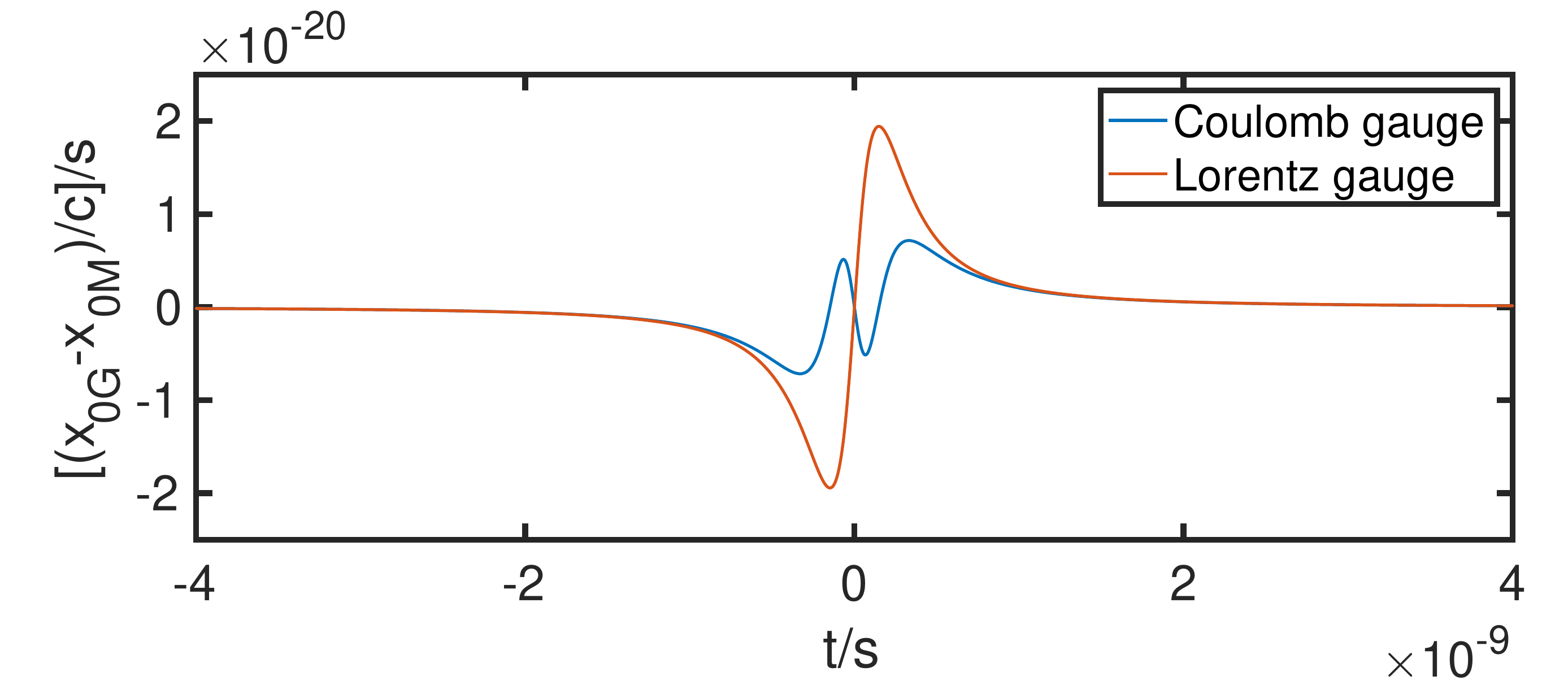}
		}
		\quad
		\subfigure[$\omega_{G}-\omega_{M}$]{
			\includegraphics[width=10cm]{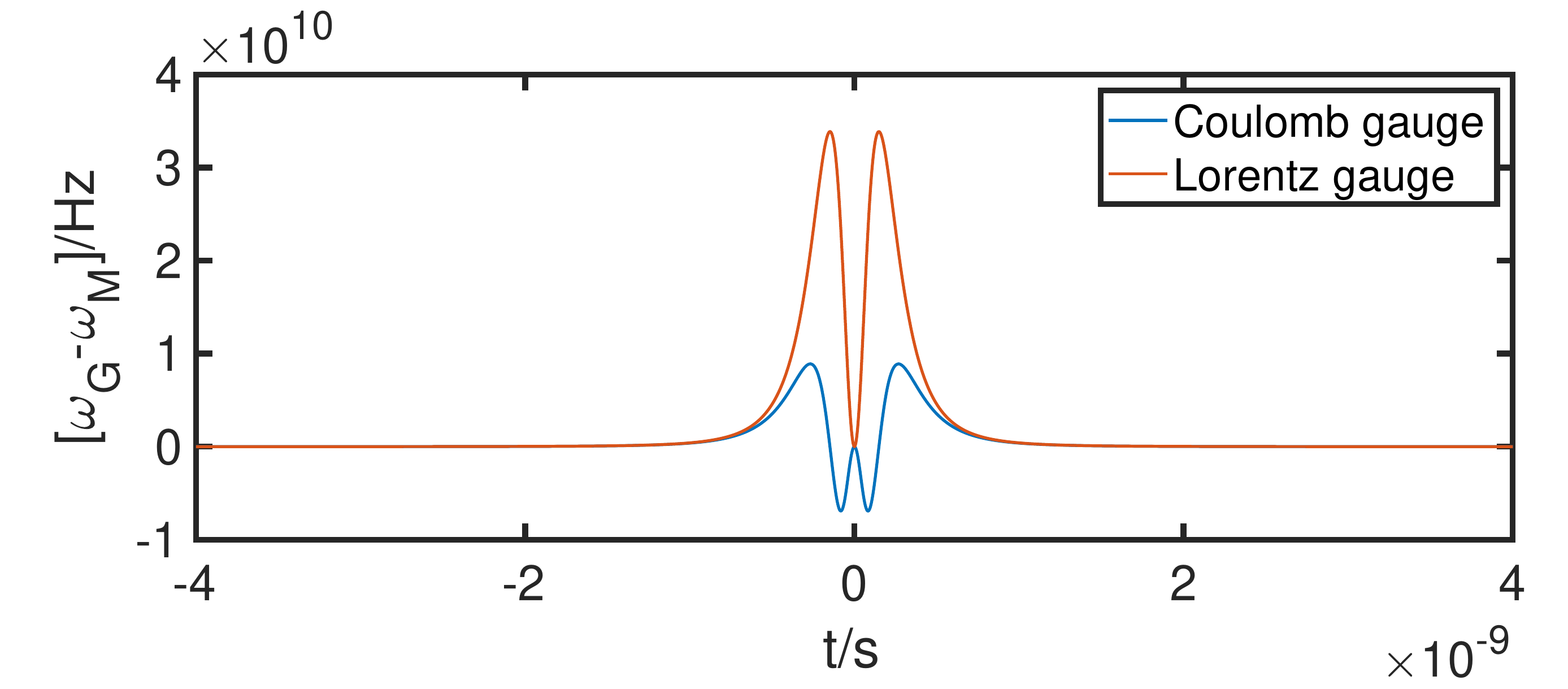}
		}
		\captionsetup{justification=raggedright}
		\caption{The reults of the instantaneous equilibrium position (a) and the frequency (b) of the oscillator in the Coulomb and Lorentz gauges, displayed as deviation from the corresponding values in the multipolar gauge.}
		\label{fig:DM}
	\end{figure}

	FIG. \ref{fig:M} gives the multipolar-gauge results of the instantaneous equilibrium position and the frequency of the oscillator. As we commented above, this is the gauge that essentially sets the equilibrium position at $\vec E=0$. FIG. \ref{fig:M1}  is an enlarged view of FIG. \ref{fig:M} for the period when the cluster acts significantly. FIG. \ref{fig:DM} gives the corresponding results in the Lorentz and Coulomb gauges, expressed as the deviation from the values in the multipolar gauge. 
	
\begin{figure}[htbp]
		\centering
		\includegraphics[width=1\linewidth]{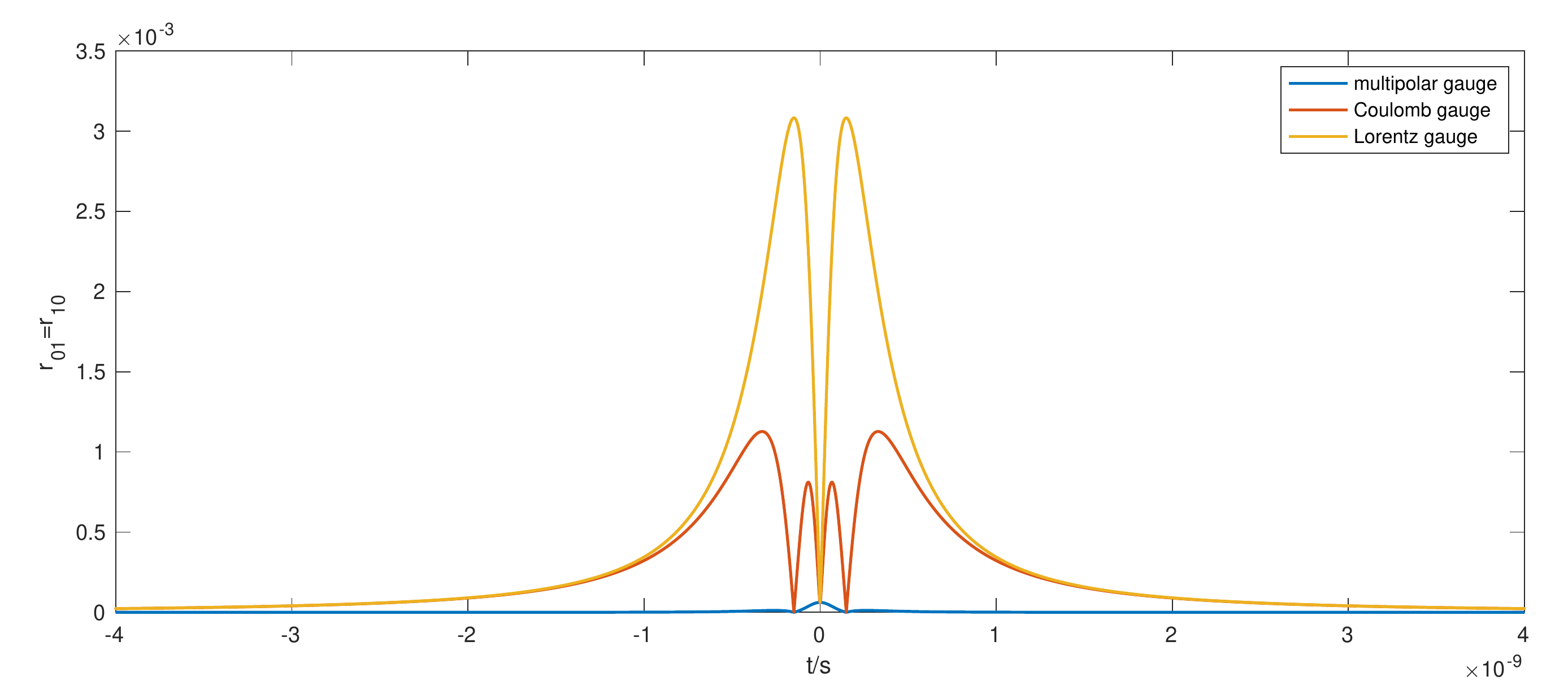}
		\captionsetup{justification=raggedright}
		\caption{ The results of the criterion parameter \(r_{01}=r_{10}\) for the adiabatic approximation in all three gauges we use.}
		\label{fig:r}
	\end{figure}

	Before solving the eigen-equations,  we must check that with our chosen parameters, $\phi(\vec x,t)$ is indeed an adiabatic potential for the oscillator, in all three gauges we use. The criterion for adiabatic approximation is the parameter
	\begin{equation}
		r_{nm}=\left|\frac{\langle n_G(t)|\dot{H}_{0G}(t)|m_G(t)\rangle}{(E_{mG}(t)-E_{nG}(t))^2}\right |,
	\end{equation}
	where $n,m$ label two different states. FIG. \ref{fig:r} gives the results of $r_{nm}$ for the lowest two states, which shows that for the Lorentz, Coulomb, and multipolar gauges, we all have  \(r_{01}=r_{10}\ll 1\). Thus, if prepared in the excited state $\left |1\right >$ before the cluster moves in, the oscillator will largely stay in the state  $\left |1_G(t)\right >$ during the whole process as the cluster moves through. 
	
	With the adiabatic conditions justified, the gauge-dependence of the instantaneous energy levels, as we just displayed above for our designed quantum oscillator, may in principle be tested experimentally. A possible observable, which is cleanest theoretically, is the spectrum of spontaneous radiation, as we will compute in the next Section.

	\section{The transient spontaneous radiation spectrum and its gauge-dependence}
	
	We now add into our system the coupling to the background vacuum electromagnetic field  \(A^\mu_B=(\vec A_{B},\phi_B)\). For the convenience of imbedding our one-dimensional oscillator into a three-dimensional formulation, we introduce an artificial harmonic potential \(V(y,z) =\frac m2 (\omega_y^2y^2+\omega_z^2z^2)\), with \(\omega_y,\omega_z\rightarrow\infty\). 
	The entire Hamiltonian is now expressed as 
	\begin{equation}{\label{HtB}}
			H(t)=\frac{1}{2m}[\vec p-q(\vec A(\vec x,t)+\vec A_{B}(\vec x,t))]^2+q(\phi(\vec x,t)+\phi_B(\vec x,t))+V(y,z)+H_B.
		\end{equation}
		Here, $H_B$ is the Hamiltonian of the background photon: 
		\begin{equation}{\label{HB}}
			H_B=\int \frac{d^3 k}{(2\pi)^3} \sum_{\lambda}\omega_{\vec{k}} a_{\vec{k},\lambda}^{\dagger} a_{\vec{k},\lambda}.
		\end{equation}
		
		The expression of $A^\mu_B$ also depends on the gauge. For the Coulomb gauge,
		\begin{equation}{\label{AB-C}}
			\vec A_{B}=\int \frac{d^3 k}{(2\pi)^3}\sum_{\lambda}\sqrt{\frac{1}{2\omega_{\vec k}}}(\vec e_{\lambda}a_{{\vec k},\lambda}e^{i\vec k\cdot\vec x-i\omega_{\vec k}t}+\vec e^*_{\lambda}a^{\dagger}_{{\vec k},\lambda}e^{-i\vec k\cdot\vec x+i\omega_{\vec k}t}),
			\quad\phi_B=0
			\quad(\vec e_{\lambda}\cdot\vec k=0).
		\end{equation}
		For the Lorentz gauge,
		\begin{equation}{\label{AB-L}}
			A_{B}^{\mu}=\int \frac{d^3 k}{(2\pi)^3} \sum_{\lambda} \sqrt{\frac{1}{2 \omega_{\vec{k}}}}\left(e_{\lambda}^{\mu} a_{\vec{k},\lambda} e^{i \vec{k} \cdot \vec{x}-i\omega_{\vec k}t}+e_{\lambda}^{\mu *} a_{\vec{k},\lambda}^{\dagger} e^{-i \vec{k} \cdot \vec{x}+i\omega_{\vec k}t}\right).
		\end{equation}
		Since we shall only consider the emission of a physical photon, and only compute this effect to leading order, in the Lorentz gauge the part of $A_{B}^{\mu}$ that really works actually agrees with Eq. (\ref{AB-C}). We therefore omit the gauge label in the background potentials.  
		
		We can roughly estimate that for our system the wavelength of the possibly detected photon is much larger than the geometric size of the oscillator, so we approximate $A_{B}^{\mu}$ as the value at the equilibrium point $x_0$ of the oscillator:  
		\begin{equation}
			\vec A_{B}\approx\int \frac{d^3 k}{(2\pi)^3}\sum_{\lambda}\sqrt{\frac{1}{2\omega_{\vec k}}}(\vec e_{\lambda}a_{\vec {k},\lambda }e^{i\vec k\cdot\vec x_{0G}(t)-i\omega_{\vec k}t}+\vec e^*_{\lambda}a^{\dagger}_{\vec {k},\lambda}e^{-i\vec k\cdot\vec x_{0G}(t)+i\omega_{\vec k}t}),
			\quad\phi_B=0\quad(\vec e_{\lambda}\cdot\vec k=0).
		\end{equation}
		Note that $x_0$ depends on the gauge for the external field, so we still need a gauge label here, \(G=C,L\). Using the PZW transformation, we can get the multipolar-gauge expression in the dipole approximation:  
		\begin{equation} 
			\vec A_{B}\approx 0\quad\phi_B\approx(\vec x-\vec x_{0M}(t))\cdot\int \frac{d^3 k}{(2\pi)^3}\sum_{\lambda}\sqrt{\frac{\omega_{\vec k}}{2}}i(\vec e_{\lambda}a_{\vec {k}, \lambda}e^{i\vec k\cdot\vec x_{0M}(t)-i\omega_{\vec k}t}-\vec e^*_{\lambda}a^{\dagger}_{\vec {k}, \lambda}e^{-i\vec k\cdot\vec x_{0M}(t)+i\omega_{\vec k}t})\quad(\vec e_{\lambda}\cdot\vec k=0).
		\end{equation}
		
		To compute the radiation spectrum, we now have two possible sources of gauge-dependence, one from the external field and the other from the background field. For the non-perturbative, adiabatic, and relativistic external field, we just explained in the previous Section that the gauge-choice seriously affects the energy levels of our quantum system. As we will shortly show, this indeed leads to significant gauge dependence of the spectrum.  For the background field, since it does not contribute a non-perturbative scalar potential, its gauge-dependence can be handled by Eq. (\ref{EA}), which agrees well with the measured atomic spectroscopy, as Lamb {\it et. al.} elaborated in Ref. \cite{Lamb87}.  
		We can expect that this method still applies when coupling the background field to our non-perturbative time-dependent system. (Although our comments in the paragraph around Eq. (\ref{Ht}) still hold here.) So, analogous to Eq. (\ref{nG}), the instantaneous eigenstate under the influence of the background electromagnetic field is defined by applying another unitary transformation:  
		\begin{equation}{\label{nBG}}
			\left| n_{BG}(t) \right> =U_{BG} \left | n_{G}(t) \right> ,
		\end{equation}
		where
		\begin{equation}
			U_{BM}= 1,\quad U_{BG}=\exp\{-ie\vec A_B(\vec x_{0G}(t),t)\cdot(\vec x-\vec x_{0G}(t))\}~(G=L,C).
		\end{equation}
		Using the above state, gauge-dependence from the background electromagnetic field is removed. Certainly, the calculation is the simplest in the multipolar gauge with $\vec A_B=0$ and $U_{BM}=1$, then $\left| n_{BG}(t) \right\rangle =\left | n_{G}(t)\right \rangle$, and the total Hamiltonian (\ref{HtB}) becomes
		\begin{equation}\label{HtM}
			H_{G}(t)\approx H_{0G}(t)+e\vec E_B(\vec x_{0G}(t),t)\cdot(\vec x-\vec x_{0G}(t))+H_B.
		\end{equation}
		
		We would like to remark that for our system the gauge-dependence from the background field is actually not serious anyhow, as compared with that from the external field. Even if we do not apply the transformation in Eq. ({\ref{nBG}), and go ahead in any gauge with $\left | n_{G}(t) \right\rangle$ and the minimal-coupling Hamiltonian
			\begin{equation}
				H(t)\approx H_{0G}(t)+\frac 1m\vec A_B(\vec x_{0G}(t),t)\cdot(\vec p+e\vec A(\vec x,t))+H_B,
			\end{equation}
			the result would differ from that using Eq. (\ref{HtM}) by an undetectable amount, as we will check at the end of this Section. 
			
			Note that our quantum oscillator is essentially one-dimensional, so we only need to consider the physical photon with \(e_x\) polarization, and the background photon Hamiltonian (\ref{HB}) reduces to 
			\begin{equation}
				H_B=\int \frac{d^3 k}{(2\pi)^3} \omega_{\vec{k}} a_{{\vec k}, x}^{\dagger} a_{\vec{k}, x} .
			\end{equation}
			Moreover, due to the constraint \(\vec e_{\lambda}\cdot\vec k=0\), the photon momentum-space integration is reduced as well:   
			\begin{equation}{\label{inte}}
				\int d^{3} k \Rightarrow
				\int_{0}^{\infty} \omega_{\vec k}^{2} d \omega_{\vec k} \int_{0}^{\pi} \sin \theta_{\vec k} d \theta_{\vec k} \delta(\theta_{\vec k}-\frac\pi2) \int_{0}^{2 \pi} d \phi_{\vec k}=2 \pi \int_{0}^{\infty} d \omega_{\vec{k}} \omega_{\vec{k}}^{2},
			\end{equation}
			where \(\theta_{\vec k}\) is the angle between the photon momentum $\vec k$ and the $x$ axis, and $\varphi _{\vec k}$ is the azimuthal angle of $\vec k$ in the $y-z$ plane. 
			
			To facilitate our discussion, we write collectively in all gauges the interaction term between the background field and the time-dependent quantum system as
			\(  \int \frac{d^3 k}{(2\pi)^3} H_I(a_{\vec {k},x},a^{\dagger}_{\vec{k},x},t)\). 
			We use energy-level raising and lowering operators \(\sigma_{+}\) and \(\sigma_{-}\) to represent the position operator \(x-x_{0G}(t)\) and the mechanical-momentum operator \(p_x+eA_x\):
			\begin{equation}
				x-x_{0G}(t)=\frac 1{\sqrt{2}\gamma_G(t)}(\sigma_++\sigma_-),
				\quad p_x+eA_x=i\frac{\gamma_G(t)}{\sqrt{2}}(\sigma_+-\sigma_-),
			\end{equation}
			where \(\sigma_+|n_G(t)\rangle=\sqrt{n+1}|(n+1)_G(t)\rangle,\sigma_-|n_G(t)\rangle=\sqrt{n}|(n-1)_G(t)\rangle\).
			By the standard practice of rotating-wave approximation, the interaction Hamiltonian can be expressed as
			\begin{equation}
				H_I(a_{\vec {k},x},a^{\dagger}_{\vec{k},x},t) \approx H_I(\sigma_+, a_{\vec {k},x},t)+H_I(\sigma_-,a^{\dagger}_{\vec{k},x},t).
			\end{equation}
			The explicit expressions,  if the multipolar gauge is employed for the background field, are   
			\begin{equation}
				\begin{aligned}
					H_I(\sigma_+, a_{\vec {k},x},t)&=\frac i2 \int \frac{d^3 k}{(2\pi)^2}\frac{\sqrt{\omega_{\vec k}}}{\gamma_G(t)}\sigma_+a_{\vec {k},x}e^{i\vec k\cdot\vec x_{0G}(t)-i\omega_{\vec k}t}\\
					H_I(\sigma_-,a^{\dagger}_{\vec{k},x},t)&=-\frac i2\int\frac{d^3 k}{(2\pi)^2}\frac{\sqrt{\omega_{\vec k}}}{\gamma_G(t)}\sigma_-a^{\dagger}_{\vec{k},x}e^{-i\vec k\cdot\vec x_{0G}(t)+i\omega_{\vec k}t}
				\end{aligned}\quad(G=M,L,C).
			\end{equation}
			Note that the gauge imprint from the external field still persists here. If the Lorentz or Coulomb gauge is employed for the background field, then
			\begin{equation}
				\begin{aligned}
					H_I(\sigma_+, a_{\vec {k},x},t)&=\frac i2\int\frac{d^3 k}{(2\pi)^2}\frac{\gamma_G(t)}{\sqrt{\omega_{\vec k}}}\sigma_+a_{\vec {k},x}e^{i\vec k\cdot\vec x_{0G}(t)-i\omega_{\vec k}t}\\
					H_I(\sigma_-,a^{\dagger}_{\vec{k},x},t)&=-\frac i2\int\frac{d^3 k}{(2\pi)^2}\frac{\gamma_G(t)}{\sqrt{\omega_{\vec k}}}\sigma_-a^{\dagger}_{\vec{k},x}e^{-i\vec k\cdot\vec x_{0G}(t)+i\omega_{\vec k}t}
				\end{aligned}\quad(G=L,C).
			\end{equation}
			
			The whole state function of the time-dependent oscillator plus the possibly emitted photon can be written as 
			\begin{equation}
				|\psi(t)\rangle=\sum_n a_n(t)|n_G(t),0\rangle+\sum_n\int \frac{d^3 k}{(2\pi)^3}a_{n,\vec k} (t)
				|n_G(t),\gamma_{\vec {k},x}\rangle+\cdots.
			\end{equation}
			Here, \(|n_G(t),0\rangle=|n_G(t)\rangle\otimes|0\rangle,|n_G(t),\gamma_{\vec {k},x}\rangle=|n_G(t)\rangle\otimes|\gamma_{\vec {k},x}\rangle\). \(|n_G(t)\rangle\) is the instantaneous eigenstate of the oscillator as we constructed in the previous Section, and \(|0\rangle,|\gamma_{\vec {k},x}\rangle\) are the Fock states of the photon. We only consider the single-photon process and the lowest two levels of the oscillator. Then, the whole state is approximated as 
			\begin{equation}
				|\psi(t)\rangle\approx\sum_{n=0,1} a_n(t)|n_G(t),0\rangle+\sum_{n=0,1}\int \frac{d^3 k}{(2\pi)^3}a_{n,\vec k}|n_G(t),\gamma_{\vec {k},x}\rangle .
			\end{equation}
			
			To handle the coefficients $a_n$ and $a_{n,\vec k}$, we introduce the dynamic phase \(\theta_n=-\int^t_0E_n(s)ds\) and adiabatic phase \(\gamma_n=i\int^t_0\langle n(s)|\dot{n}(s)\rangle\):
			\begin{equation}
				\begin{aligned}
					a_n(t)&=\exp[i(\gamma_n(t)+\theta_n(t))]c_n(t),\\
					a_{n,\vec k}(t)&=\exp[i(\gamma_n(t)+\theta_n(t))]c_{n,\vec k}(t) .
				\end{aligned}
			\end{equation}
			Using the state-evolution equation \(i\partial_t|\psi(t)\rangle=H(t)|\psi(t)\rangle\), we obtain differential equations for the new coefficients $c_n$ and $c_{n,\vec k}$:
			\begin{equation}{\label{c}}
				\begin{aligned}
					\dot{c}_0(t)&=-\langle 0_G(t),0|\dot{1}_G(t),0\rangle c_1(t)\exp[i(\gamma_1(t)+\theta_1(t))-i(\gamma_0(t)+\theta_0(t))],\\
					\dot{c}_{0,\vec k}(t)&=-\int\frac{d^3k'}{(2\pi)^3}\langle 0_G(t),\gamma_{\vec {k},x}|\dot{1}_G(t),\gamma_{\vec {k'},x}\rangle c_{1,\vec k'}(t)\exp[i(\gamma_1(t)+\theta_1(t))-i(\gamma_0(t)+\theta_0(t))]\\
					&-i\langle 0_G(t),\gamma_{\vec {k},x}|H_I(\sigma_-,a_{\vec {k},x}^{\dagger},t)|1_G(t),0\rangle c_1(t)\exp[i(\gamma_1(t)+\theta_1(t))-i(\gamma_0(t)+\theta_0(t))],\\
					\dot{c}_1(t)&=-\langle 1_G(t),0|\dot{0}_G(t),0\rangle c_0(t)\exp[-i(\gamma_1(t)+\theta_1(t))+i(\gamma_0(t)+\theta_0(t))]\\
					&-i\int \frac{d^3 k}{(2\pi)^3}\langle 1_G(t),0|H_I(\sigma_+,a_{\vec {k},x},t)|0_G(t),
					\gamma_{\vec {k},x}\rangle c_{0,\vec k}(t)\exp[-i(\gamma_1(t)+\theta_1(t))+i(\gamma_0(t)+\theta_0(t))],\\
					\dot{c}_{1,\vec k}(t)&=-\int\frac{d^3k'}{(2\pi)^3}\langle 1_G(t),\gamma_{\vec {k},x}|\dot{0}_G(t),\gamma_{\vec {k'},x}\rangle c_{0,\vec k'}(t)\exp[-i(\gamma_1(t)+\theta_1(t))+i(\gamma_0(t)+\theta_0(t))].
				\end{aligned}
			\end{equation}
			The notations are \(\langle n_G(t),0|\dot{m}_G(t),0\rangle=\langle n_G(t)|\otimes\langle 0|\partial_t|m_G(t)\rangle\otimes|0\rangle=\langle n_G(t)|\partial_t|m_G(t)\rangle\otimes\langle 0|0\rangle=\langle n_G(t)|\partial_t|m_G(t)\rangle\otimes\mathbf{I}\) and \(\langle n_G(t),\gamma_{\vec k,x}|\dot{m}_G(t),\gamma_{\vec k',x}\rangle=\langle n_G(t)|\otimes\langle \gamma_{\vec k,x}|\partial_t|m_G(t)\rangle\otimes|\gamma_{\vec k',x}\rangle=\langle n_G(t)|\partial_t|m_G(t)\rangle\otimes\langle \gamma_{\vec k,x}|\gamma_{\vec k',x}\rangle=(2\pi^3)\delta^{(3)}(\vec k-\vec k')\langle n_G(t)|\partial_t|m_G(t)\rangle\otimes\mathbf{I}\) . 
			It would be very complicated to calculate the complete evolution process. Fortunately, we demonstrated in Section II that the moving charge cluster acts adiabatically, therefore the transition terms which conserve the photon number can be neglected, and we only need to consider the emission and absorption of photon caused by coupling with the background electromagnetic field. Then Eqs.  (\ref{c}) simplify greatly to  
			\begin{equation}\label{c01}
				\begin{aligned}
					\dot{c}_{0,\vec k}(t)&=-i\langle 0_G(t),\gamma_{\vec {k},x}|H_I(\sigma_-,a_{\vec{k},x}^{\dagger},t)|1_G(t),0\rangle c_1(t)\exp[i(\gamma_1(t)+\theta_1(t))-i(\gamma_0(t)+\theta_0(t))],\\
					\dot{c}_1(t)&=-i\int \frac{d^3 k}{(2\pi)^3}\langle 1_G(t),0|H_I(\sigma_+,a_{\vec {k},x},t)|0_G(t),\gamma_{\vec {k},x}\rangle c_{0,\vec k}(t)\exp[-i(\gamma_1(t)+\theta_1(t))+i(\gamma_0(t)+\theta_0(t))].
				\end{aligned}
			\end{equation}
			
			To handle the decay of the excited state,  we follow a method similar to the Weisskopf-Wigner approximation \cite{Weisskopf30}: 
			\begin{equation}{\label{WW}}
				\begin{aligned}
					\dot{c}_1(t)&=-\int \frac{d^3 k}{(2\pi)^3}\langle 1_G(t),0|H_I(\sigma_+,a_{\vec{k},x},t)|0_G(t),\gamma_{\vec {k},x}\rangle \exp[-i(\gamma_1(t)+\theta_1(t))+i(\gamma_0(t)+\theta_0(t))]\\
					&\times\int^t_0dt'\langle 0_G(t'),\gamma_{\vec {k},x}|H_I(\sigma_-,a_{\vec{k},x}^{\dagger},t')|1_G(t'),0\rangle c_1(t')\exp[i(\gamma_1(t')+\theta_1(t'))-i(\gamma_0(t')+\theta_0(t'))].
				\end{aligned}
			\end{equation} 
			
			To proceed with the computation in a clearer form, we denote \(\Delta_G(t)=E_{1G}(t)-E_{0G}(t)-i\langle 1_G(t)|\dot{1}_G(t)+ i\langle 0_G(t)|\dot{0}_G(t)\rangle\).
			Since the emission spectrum typically has a peak frequency, we can effectively perform the frequency integration in the range  \((-\infty, +\infty)\), and get 
			\begin{equation}{\label{delta}}
				\int_{-\infty}^{\infty} d \omega_{k} e^{i\left(\omega-\omega_{\vec k}\right)\left(t-t^{\prime}\right)}=2 \pi \delta\left(t-t^{\prime}\right).
			\end{equation}
			
			Using Eqs. (\ref{inte}), (\ref{WW}) and (\ref{delta}), we obtain
			\begin{equation}\label{c1}
				\dot{c}_1(t)=-\Gamma_G(t)c_1(t),
			\end{equation}
			where the time-dependent decaying rate is,
			\begin{equation}
				\begin{aligned}
					\Gamma_G(t)&=\frac{\Delta^2_G(t)}{2\pi}\langle 1_G(t),0|H_I(\sigma_+,a_{\Delta_G(t),x},t)e^{i\Delta_G(t)t}|0_G(t),\gamma_{\Delta_G(t),x}\rangle\\
					&\times\langle 0_G(t),\gamma_{\Delta_G(t),x}|H_I(\sigma_-,a^{\dagger}_{\Delta_G(t),x},t)e^{-i\Delta_G(t)t}|1_G(t),0\rangle.
				\end{aligned}
			\end{equation}
			
			Then, to obtain the solutions for $c_1(t)$ and $c_{0,\vec k}(t)$, we just combine Eq. (\ref{c1}) with the first line of Eqs. (\ref{c01}), compute the relevant matrix element, and perform the numerical integration step by step. Finally, the observed emission spectrum, accumulated till a time $t_f$, is computed via: 
			\begin{equation}{\label{s}}
				\int d\omega_{\vec k} S(\omega_{\vec k},t_f)=\int \frac{d^3 k}{(2\pi)^3}|c_{0,\vec k}(t_f)|^2
				\delta(\theta _{\vec k}-\frac{\pi}{2}) 
				=\int d\omega_{\vec k} \frac{\omega_{\vec k}^2}{(2\pi)^3}
				\int d \theta_{\vec k} d\varphi_{\vec k} \sin \theta_{\vec k} |c_{0,\vec k}(t_f)|^2
				\delta(\theta _{\vec k}-\frac{\pi}{2}) ,
			\end{equation}
			Explicitly,  
			\begin{equation}{\label{s1}}
				S(\omega_{\vec k},t_f) =\frac{\omega_{\vec k}^2}{(2\pi)^2}
				|c_{0,\vec k}(t_f)|^2, ~~~ ({\theta _{\vec k}=\frac{\pi}{2}}).
			\end{equation}
			
			FIG. \ref{fig:s1} gives our calculated results of the spontaneous-emission spectrum $S(\omega_{\vec k},t_f)$, and FIG. \ref{fig:ls} shows an enlarged view around the peak frequency. 
			
			\begin{figure}[htbp]
				\centering
				\includegraphics[width=1\linewidth]{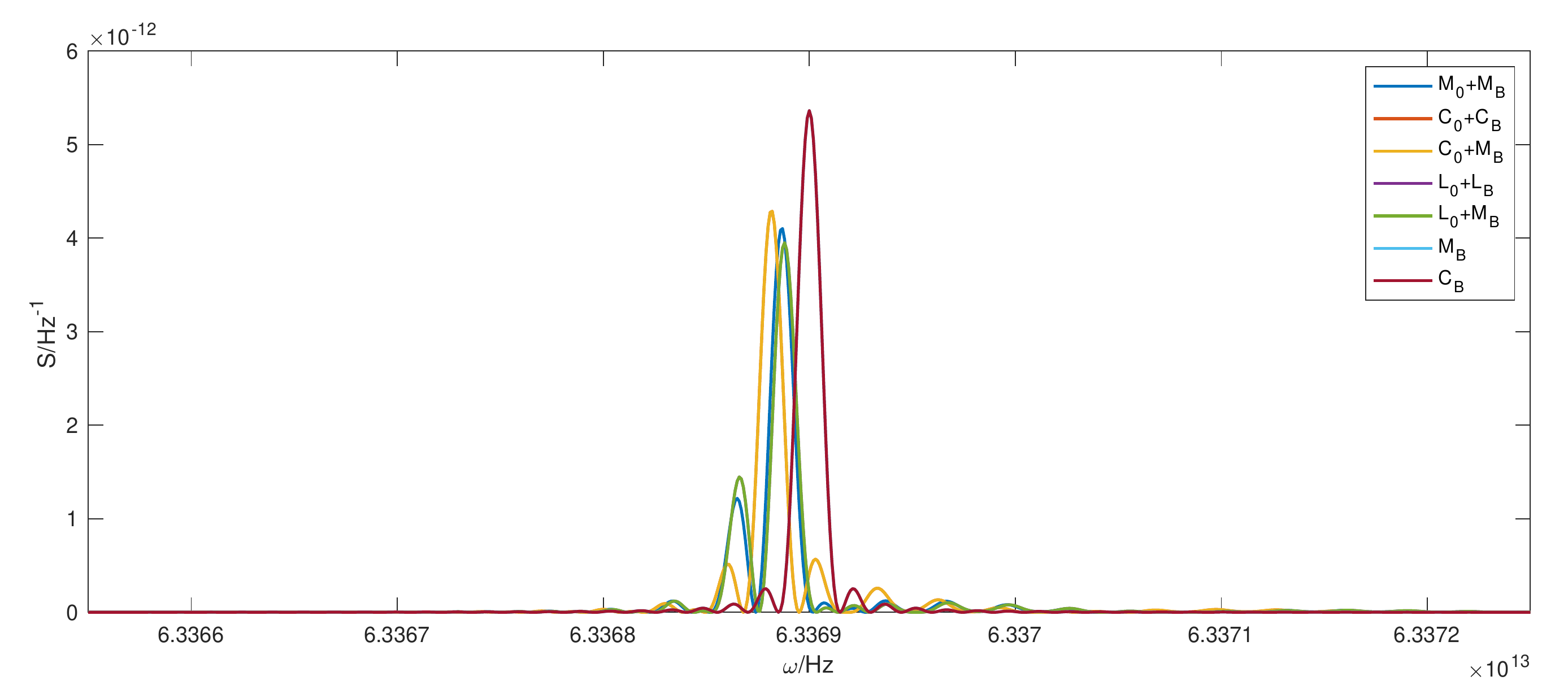}
				\captionsetup{justification=raggedright}
				\caption{The transient spontaneous-emission spectrum of our physical system. The subscript ``\(0\)'' represents the gauge selection of the electromagnetic field produced by a cluster of relativistic charged particles, and the subscript ``\(B\)'' represents the gauge selection of the background vacuum electromagnetic field. The last two lines in the legend represent the unperturbed spectrum in the absence of the charge cluster calculated in two different gauge selections of the background vacuum electromagnetic field.}
				\label{fig:s1}
			\end{figure}
			
			\begin{figure}[htbp]
				\centering
				\includegraphics[width=1\linewidth]{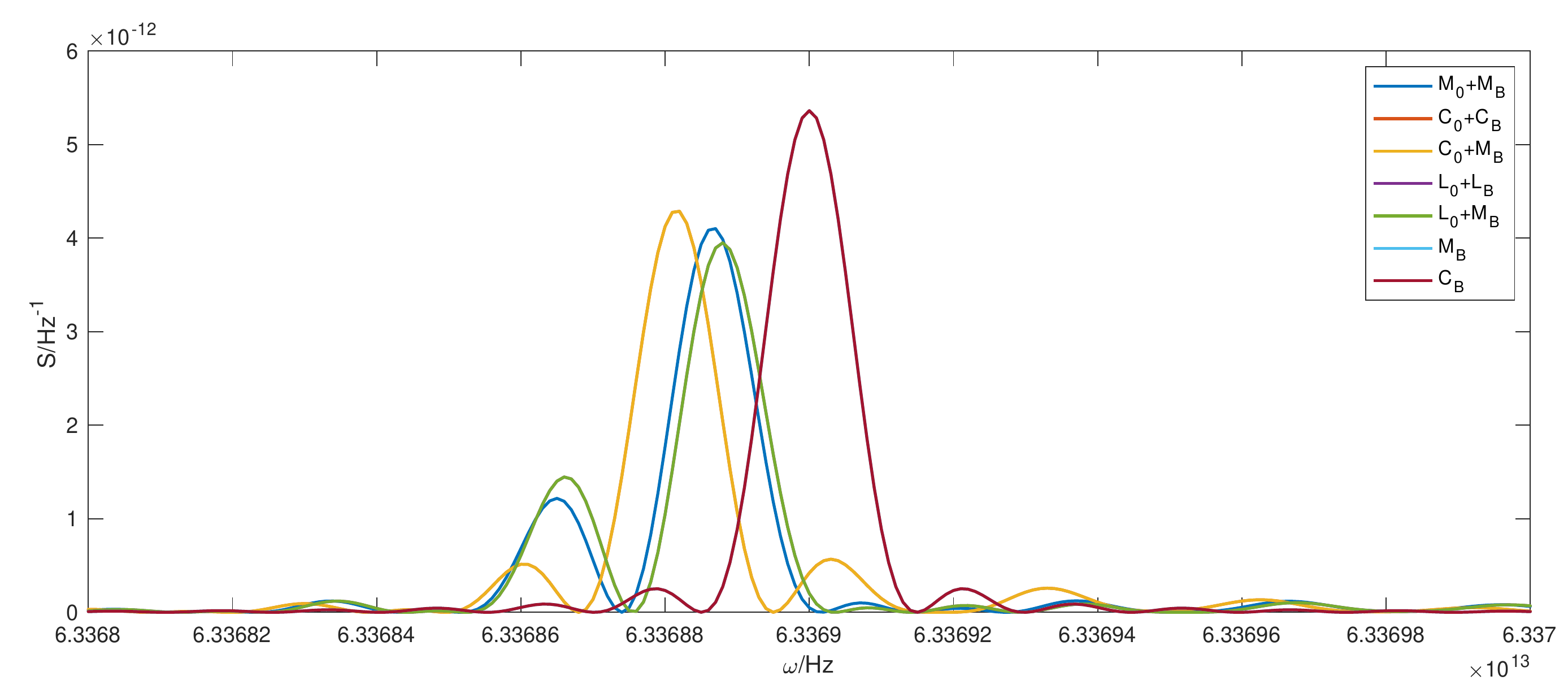}
				\caption{The enlarged view of FIG. \ref{fig:s1} around the peak frequency.}
				\label{fig:ls}
			\end{figure}
			
			These results show clearly that the computed spontaneous-emission spectrum depends considerably on the gauge, in both the peak frequency and the emission intensity. Even with a relatively mild velocity of $\beta=0.1$ for the charge cluster, the computed peak frequency in the Lorentz gauge is higher than that in the multipolar gauge by a detectable amount $\sim 10 \mathrm{M Hz}$, and higher than that in the Coulomb gauge by $\sim 60\mathrm{M Hz}$. 
			
			For another comparison, we also plot in FIG. \ref{fig:s1} and FIG. \ref{fig:ls} the unperturbed spectrum in the absence of the charge cluster, whose peak frequency is around $6.3369\times 10^{13}\mathrm{Hz}$, and is higher than that of the perturbed spectrum in the Lorentz gauge by about $120\mathrm{M Hz}$. This is a cross-check that the charge cluster indeed acts non-perturbatively on the quantum oscillator. 
			
			To close this Section, we would like to comment that the significant gauge-dependence we revealed above comes mainly from the non-perturbative, adiabatic scalar potential produced by the relativistic charge cluster. In comparison, the gauge-dependence similar to that as Lamb originally noticed  \cite{Lamb52}, is quantitatively negligible in our case, as we show in FIG. \ref{fig:DS1}.  
			\begin{figure}[htbp]
				\centering
				\includegraphics[width=1\linewidth]{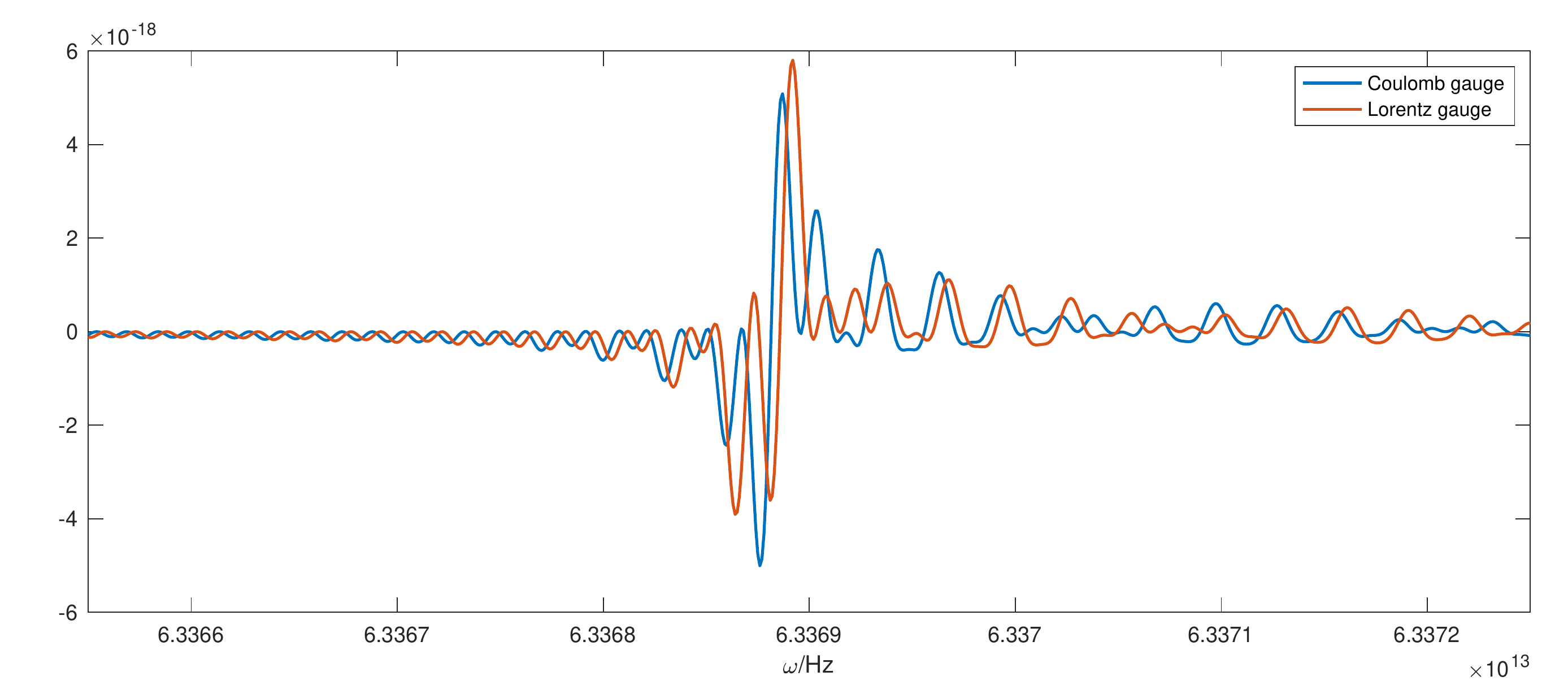}
				\captionsetup{justification=raggedright}
				\caption{The ``Lamb-type'' gauge-dependence, displayed as the spectrum deviation from the multipolar-gauge result.}
				\label{fig:DS1}
			\end{figure}
		
		\section{Summary and discussion}
		
		In this paper, we proposed a physical system that a cluster of relativistic charges produces a non-perturbative Coulomb field which acts adiabatically on a quantum system. This is a case that we have to face sharply the long-standing concerns about what is actually the physical significance of the scalar potential when it is time-dependent, and how the electromagnetic field actually interacts with a quantum system, especially, how to quantize a bound state in the  presence of a time-dependent, non-perturbative, and adiabatic scalar potential. We calculated the spontaneous-radiation spectrum of such a quantum system, and find significant gauge-dependence which cannot be cured by existing methods. 
		
		Whenever some gauge-dependence stands in the way, there is often an intent and attempt to discard $(\vec A, \phi)$ and stick to $(\vec E, \vec B)$, especially in the early practices and debates. However, a consensus was gradually reached that this is not possible, at least in a local formulation.  For example, the multipolar-gauge expressions involve only $(\vec E, \vec B)$, but in terms of path-integrals. It should especially be reminded that such integrals are in general path-dependent. Namely, the multipolar gauge itself is not unique. 
		
		Another intent of long history is to distinguish between energy and hamiltonian for a non-conservative system, and refuse assigning the time-dependent $\phi(\vec x,t)$ any serious physical significance. We do not claim against such an intent, and just bring a case that the existing methods have to utilize $\phi(\vec x,t)$, with a gauge-dependent result. 
		
		There also exists an intent in the opposite way, especially after the work of Aharanov and Bohm \cite{AB}, to think of $(\vec A, \phi)$ as physical reality. Another support for this idea is that the Coulomb field always accompanies a charge, and can never be stripped away.  We do not claim either that our findings add to that intent. In fact, we must be clear-minded that our starting point, namely the semi-classical Hamiltonian (\ref{Ht}) or (\ref{HtB}) under the external-field approximation, is known to be unsafe if the external source is relativistic \cite{Wein95}, while a safe and fully quantum-field method is still lacking. In such a circumstance, the gauge-dependence we just found is not a disaster, but an advantage: Since the Hamiltonian (\ref{Ht}) or (\ref{HtB}) is not justified anyway, there is no sense in sticking to the actual $(\vec A, \phi)$ generated by the external source. Instead, one may always try to define an {\it effective} external field which, when plugged into the Hamiltonian (\ref{Ht}) or (\ref{HtB}), may fit well the experimental data. This is indeed quite probable, as we saw that by just tuning the gauge, we may fit the peak frequency of the radiation spectrum; and another overall factor may be multiplied to fit the radiation intensity. Namely, the {\it effective} external field may likely be parameterized with the gauge potential in a particular gauge. This gauge is not necessarily among the most common ones, and might possibly be individualized condition for the field of each charge, instead of the usual conditions for the overall field.  
		
		Such an effective external field is not merely of important phenomenological use, but might also shed light on the physical significance of the gauge potential. Should it turn out to be closer to the multipolar-gauge expression, it might suggest that $\vec E$ is indeed more essential than $\phi$. Should it turn out to be closer to the expression of the Lorentz or Coulomb  gauge, and deny that of the multipolar gauge, however, we would find an extremely interesting phenomenon which may be termed ``kinematic scalar Aharanov-Bohm effect'': Due to Lorentz contraction, a charged particle close to the speed of light produces negligible electric field along its trajectory. However, its scalar potential could possibly be felt by a quantum system. We may even put this in a science-fiction style: The classical bodies can never feel a projectile flying directly to us at nearly the speed of light, no matter how crazily energetic, until the (too-late) moment it crashes in; while a quantum system may act as an alerting whistle. Pitifully, we human beings may belong to the ``classical bodies''. Such a projectile might be a particle, or even a planet. Should we expect that gravity display a similar effect, then even a massive black hole flying directly to us at nearly the speed of light can only be detected beforehand by a quantum system, but not by classical bodies.
		
		We should compare our physical system with related phenomena. The key elements of our design include: a relativistic charge cluster of a huge number, a quantum oscillator of relatively large spatial extension and low frequency, and adiabatic conditions for the oscillator to complete transient spontaneous radiation during the interacting period. These elements have been encountered individually in some studies. For example, Coulomb excitation \cite{Bayman05,Dasso06} may also utilize a relativistic projectile, but in a perturbative and/or non-adiabatic way. The AC-stark\cite{Autler55} effect may be adiabatic and non-perturbative, but the electric field is from an electromagnetic wave, not relativistic particles. The term ``strong-field'' typically refers to an intensive laser field, not a relativistic Coulomb field. It is our study that combines all these key elements, and brings sharply a ``Coulomb-type'' gauge dependence. We invent this terminology to refer to the gauge-choice problem for the Coulomb field of charged particles, especially relativistic ones. Naturally, the gauge-choice problem for an electromagnetic wave or the vacuum field may be termed ``Lamb-type''.  
		
		Since our system has  very low energy, the decaying rate is rather small. This also reveals clearly that the gauge-dependence we encounter comes mainly from defining the quantum basis, and is not the Lamb-type. To further address this new type of gauge dependence, other quantum systems of large spatial extension may also be employed, especially the fancy Rydberg atoms, for which the Coulomb field of a relativistic charged cluster may become non-perturbative relatively easily.  
		
		In this paper, we displayed gauge-dependence for the accumulated radiation spectrum without time-resolution. Registering accurately the arrival time of the emitted photons would reveal more detailed information. If time-resolution is applied, one may also consider induced transitions, either absorption or emission. 
		
		The design we proposed also makes a tunable system to study the relativistic bound-state problems. The usual relativistic bound states often go to two extremes: either all particles are relativistic and the problem got over complicated, or, the acted particle is relativistic while the acting particles non-relativistic, which can be easily handled by the Dirac equation.  In our design, the acting cluster is relativistic, while the acted particle is non-relativistic.  (One may feel like to turn the situation around by going to the rest frame of the acting cluster. But this idea would not work, since we may easily arrange two or more acting clusters with different velocities.) Moreover, the design makes both the relativity parameter $\beta$ and the interaction strength continuously adjustable.  
		
		The ultimate goal of this study is to thoroughly examine how the gauge field actually interacts with a quantum system, and explore more closely the physical significance of the gauge potential. A similar system as we proposed may also be designed to address gravitational gauge dependence, by studying quantum transitions in the presence of a time-dependent relativistic non-perturbative Newtonian field. 
		\newpage
			
			\section{Acknowledgements}
			
			This is a long-time work that almost all our group members have joined the discussion, including our graduated fellows. We also benefited from fruitful discussions with many HUST colleagues, especially Jian-Wei Cui, Wei-Tian Deng, Lin Li, and Yi-Qiu Ma, et al.  The work had been, and was partly supported by the China NSF via Grants No. 11275077 and No. 11535005.


\begin{thebibliography}{99}
				
				\bibitem{Lamb52}W. E. Lamb, Phys. Rev. \textbf{85}, 259 (1952)
				
				\bibitem{Yang76}K. H. Yang, Ann. Phys. \textbf{101}, 62 (1976)
				
				\bibitem{Yang81}K. H. Yang, Phys. Lett. A \textbf{84}, 165 (1981)
				
				\bibitem{Au84}C. K. Au, J. Phys. B: At. Mol. Opt. Phys, \textbf{17}, L59 (1984)
				
				\bibitem{Lamb87}W. E. Lamb, R. R. Schlicher, and M. O. Scully,
				Phys. Rev. A \textbf{36}, 2763 (1987)
				
				\bibitem{Funai19} N. Funai, J. Louko, and E. Mart\'{\i}n-Mart\'{\i}nez, Phys. Rev. D \textbf{99} 065014 (2019)
				
				\bibitem{Note} Note that including the total $H(t)$ to define the instantaneous eigenstates, thus   the interaction term is apparently absent, does not mean that the system will not make quantum transition. The reason is that $H(t)$ may not commute at different times, therefore the eigenstate at one moment is not always the eigenstate later. As is known in the discussion of adiabatic approximation, quantum transitions are avoided only if $H(t)$ varies slowly enough in time.
				
				\bibitem{Power59}E. A. Power and S. Zienau. Phil. Trans. R . Soc. Lond. A \textbf{251}, 427 (1959)
				
				\bibitem{Wolley71}R. G. Woolley, Proc. R . Soc. Lond. A \textbf{321}, 557 (1971)
				
				\bibitem{Weisskopf30}V. F. Weisskopf and E. P. Wigner, Z. Phys. \textbf{63}, 54 (1930)
				
				\bibitem{AB}  Y. Aharonov and D. Bohm, Phys. Rev. \textbf{115}, 485 (1959).
				
				\bibitem{Wein95}  See, for example, S. Weinberg, The Quantum Theory of Fields Vol. I (Cambridge, New York,
				1995), section 13.6.
				
				\bibitem{Bayman05}B. F. Bayman and F. Zardi, Phys. Rev. C \textbf{71}, 014904 (2005)
				
				\bibitem{Dasso06}C. H. Dasso, M.I. Gallardo, H. M. Sofia and A. Vitturi,  Phys. Rev. C \textbf{73}, 034612 (2006)
				
				\bibitem{Autler55}S. H. Aulter and C. H. Townes, Phys. Rev. \textbf{100}, 703 (1955)
				
				
			\end{thebibliography}
		\end{document}